\documentclass{article}

\usepackage{arxiv}

\usepackage[utf8]{inputenc} % allow utf-8 input
\usepackage[T1]{fontenc}    % use 8-bit T1 fonts
\usepackage{hyperref}       % hyperlinks
\usepackage{url}            % simple URL typesetting
\usepackage{booktabs}       % professional-quality tables
\usepackage{amsfonts}       % blackboard math symbols
\usepackage{amsmath}        % for \dfrac and advanced math
\usepackage{nicefrac}       % compact symbols for 1/2, etc.
\usepackage{microtype}      % microtypography
\usepackage{lipsum}		% Can be removed after putting your text content
\usepackage{ulem}
\usepackage{graphicx}
\usepackage{natbib}
\usepackage{doi}
\usepackage{xcolor}

\title{Are NICER and GW170817 constraints suggesting a compactified scenario for Neutron stars?}
\author{ \href{https://orcid.org/0009-0000-8375-4833}{\includegraphics[scale=0.06]{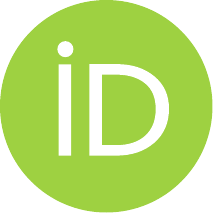}\hspace{1mm}}Asim Kumar Saha\thanks{E-mail: asim21@iiserb.ac.in} \\
Indian Institute of Science Education and Research Bhopal\\
	\And
	\href{https://orcid.org/0000-0003-2633-5821}{\includegraphics[scale=0.06]{orcid.pdf}\hspace{1mm}} Tuhin Malik\thanks{E-mail: tuhin.malik@uc.pt}\\
	Departamento de Física, Universidade de Coimbra, 3004-516 Coimbra, Portugal \\
    \And
	\href{https://orcid.org/0000-0003-2943-6388}{\includegraphics[scale=0.06]{orcid.pdf}\hspace{1mm}} Ritam Mallick\thanks{E-mail: mallick@iiserb.ac.in} 
    \\
	 Indian Institute of Science Education and Research Bhopal\\}
\begin{document}
\label{firstpage}
%\pagerange{\pageref{firstpage}--\pageref{lastpage}}
\maketitle

\begin{abstract}
Astrophysical observations from NICER and gravitational wave data constrain the properties of matter at the cores of neutron stars, enabling us to probe high-density matter with greater accuracy. To understand its implications for neutron stars, three distinct class-agnostic equation-of-state ensembles are constructed using the speed-of-sound parametrisation, which can describe matter in neutron-star cores. Bayesian analysis is employed to constrain the parameters—namely, the squared speed of sound and chemical potential using the observational data. The Bayesian inference shows that the observations effectively constrain the low-density region of the equation of state. The astrophysical bound favours a softer, low-density equation of state in which the phase transition occurs at intermediate densities, thereby reducing the upper mass bounds for neutron stars. For the equation of state with density discontinuity, the discontinuities are preferably small. The equation of state with maximum mass configuration shows considerable stiffening from very low density, providing pressure support to generate maximum mass. In contrast, the equation of state with the maximum compact stellar configuration has a softer low-density equation of state, followed by pronounced stiffening, yielding the maximum compact configuration. The observationally favoured EoS shares the same qualitative structure as the maximum-compactness EoS: relative softness at intermediate densities transitioning to stiffness at high densities, a configuration gravity naturally favours.
\end{abstract}

% \maketitle
%\flushbottom


\section{Introduction}
The equation of state of matter at densities exceeding those found in atomic nuclei remains one of the most profound unsolved problems in physics. The cores of neutron stars (NSs) are incredibly dense, where matter exists under extreme gravitational conditions \citep{ShapiroTeukolsky1983, Oertel2017, LattimerPrakash2007, Glendenning2000}. Such a captivating system has therefore long intrigued physicists. The scenario gains additional importance as ab initio theoretical and experimental advances to understand the matter at such densities have proved challenging until now \citep{de-forcrand, goy, lattimer2004}. Therefore, the only natural laboratories for high-density matter are the cores of the NSs \citep{Glendenning2000, LattimerPrakash2007, OzelFreire2016}. For the last few decades, physicists have sought to reconcile model predictions of NSs with observations to understand matter at extreme density and gravity \citep{Huth:2020ozf, Huth:2021bsp, Annala:2021gom, Nicholl:2021rcr, Imam:2024gfh, Biswas:2024hja}. At the heart of the problem lies the quest to apprehend the unresolved regime of quantum chromodynamics (QCD), the state of matter at high density and low temperature. The uncertainties of the finite density regime arise from the ignorance of matter properties beyond nuclear saturation density. The chiral effective field theory ($\chi$ EFT) models the sub-nuclear densities satisfactorily but is not equipped to extend beyond such densities \citep{Hebeler_2013, Gandolfi:2019zpj, Keller:2020qhx}, whereas the asymptotic high-density matter regime is approximately captured with perturbative QCD calculations \citep{Kurkela2010pQCD, Fraga:2013qra, Kurkela_2014}.

Over the last decade or two, observational astrophysics has pushed the boundaries, yielding precise observations of NSs, including their periods, masses, radii, and tidal deformabilities, among other properties \citep{Antoniodis_2013, Cromartie_2020, Fonseca_2021, GW170817, Abbott_2019, Miller_2019, Riley_2019, Miller_2021, Riley_2021}. The gravitational waves (GW) inference of the tidal deformability and Neutron Star Interior Composition Explorer (NICER) inference of the mass and radius for a few pulsars have provided stringent constraints on the equation of state (EoS) of matter that resides at the cores of the NSs \citep{GW170817, Abbott_2019, Miller_2019, Riley_2019, Miller_2021, Riley_2021}. Along with these observations, agnostic approaches that depend only on the thermodynamic consistency of matter (as opposed to detailed microphysics, which are highly model-dependent) have proven very effective in constraining the EoS at densities achieved by NSs \citep{Read2009Polytrope, Lindblom2010Spectral, Greif:2018njt, LandryEssick2019GP, Annala2020, Somasundaram2021MuEoS}. The agnostic approaches also rely on limiting conditions at low density and at asymptotic high density, where chiral effective field theory \citep{Hebeler_2013, Gandolfi:2019zpj, Keller:2020qhx} and perturbative calculations \citep{Kurkela2010pQCD, Fraga:2013qra, Kurkela_2014} are trusted with relatively high confidence.

The collaborative effort of NICER/GW170817 observations and an agnostic approach has significantly constrained the possible EoS of high-density matter \citep{Annala2020, Altiparmak:2022bke, kamal_2023, pratik2024, ecker_2022, ecker_2023}. However, the understanding of high-density matter is far from complete. Presently, one cannot even distinguish between the distinctly different EoS that can be present at NS cores \citep{gorda, Verma:2025vkk, Verma:2025dez}. The EoS can be broadly categorised into three classes according to how the speed of sound depends on density \citep{Verma:2025dez}. The speed of sound carries the information of the EoS by definition ($c_s^2=\frac{\partial p}{\partial \epsilon}$, where $p$ is the pressure and $\epsilon$ is the energy density) \citep{HaenselPotekhinYakovlev2007, RezzollaZanotti2013, Raduta_2022, Altiparmak:2022bke, Roy:2022nwy}. One can classify them as monotonically increasing (usually characterised by hadronic EoS), non-monotonic (usually associated with smooth phase transition (PT)) and discontinuous (associated with strong PT) EoS. Although they are distinctly different, the NICER/GW170817 observation still does not favour one over the other \citep{Verma:2025dez}. A fundamental question, therefore, remains: do these observational constraints reveal any universal pattern in the structure of dense matter? Additionally, what exactly is the role of gravity in defining the NS, along with the EoS?

Here we address these questions through a comprehensive Bayesian analysis of three distinct classes of agnostic EoS ensembles, constrained by the complete set of available NICER observations---including the recently reported PSR~J0614$-$3329 \citep{Mauviard:2025dmd}---and gravitational wave data from GW170817. Unlike the usual rejection-sampling methods \citep{Altiparmak:2022bke, Annala2020}, our approach directly explores the posterior distribution, yielding more efficient, probabilistically rigorous inference of the EoS parameters. The maximum mass and maximum compactness sequences have also been studied in detail to scan for any pattern evolving from the astrophysical constraints \citep{Rhodes, lindbolm, annurev_Lattimer, Rezzolla_2018, Margalit_2017, rocha2023, musolino2023, irfan, anshuman, Gao:2025vdc, Imam:2025lut, Rezzolla:2025pft}.

Our analysis reveals a striking pattern. We compare three characteristic EoS from our ensemble: the one yielding maximum mass, the one yielding maximum compactness, and the one most favoured by observations. The maximum mass EoS is stiff at intermediate densities---this early stiffening provides the pressure support needed to sustain the largest stellar masses. However, the most probable EoS, as constrained by NICER and GW170817, is soft at intermediate densities and stiffens only at high densities. Crucially, this soft-to-stiff structure mirrors the maximum compactness EoS, not the maximum mass EoS. Current observations are therefore selecting for EoS that produce compact stars rather than massive stars. This suggests a fundamental property of compact objects: the compactification of NSs, a universal property of gravity that compresses and compactifies objects.

\section{Formalism}
\paragraph{EoS Construction}
Tabulated BPS EoS \citep{BPS} defines the low-density matter till approximately $0.5 ~n_0$, \textbf{$n_0 \approx 0.16 ~\mathrm{fm}^{-3}$} is the nuclear saturation density. A polytrope of the form $P = K n^{\Gamma}$ defines matter till $1.1 ~n_0$, where the adiabatic index $\Gamma$ is varied within the range $[1.77,3.23]$ to span the CET uncertainty band \citep{Hebeler_2013, Altiparmak:2022bke}. Beyond this density, the speed-of-sound interpolation formulation is used to define the high-density matter, where the squared speed of sound, $c_s^2$, is parametrised as a function of chemical potential $\mu$. The number density and pressure are then computed from these two parameters. The implementation uses piecewise-linear interpolation of $c_s^2(\mu)$ defined over four randomized segments, specified by randomised points $(\mu_i, c_{s,i}^2)$, where $\mu_i \in [\mu_{\scriptscriptstyle CET}, 2.6 ~\mathrm{GeV}]$. This upper limit on $\mu$ follows the prescription of \citep{Fraga:2013qra}, chosen such that the uncertainty in the perturbative QCD (pQCD) regime matches that in the CET region. 

The study focuses on three broad classes of EoS: monotonic, non-monotonic, and discontinuous, as outlined in \citep{Verma:2025dez}. The key criterion for this classification is the behaviour of the speed of sound ($c_s^2$) within the star, particularly the location of its peak. For each generated EoS, the corresponding maximum mass configuration ($M_{\rm TOV}$) is determined. The $M_{\rm TOV}$ denotes the gravitational mass obtained by solving the Tolman--Oppenheimer--Volkoff (TOV) equations \citep{tov}. If the speed of sound decreases monotonically with increasing radius in the $M_{\rm TOV}$ star profile, the EoS is classified as monotonic. If the maximum value of the speed of sound occurs away from the centre of the star, the EoS is labelled non-monotonic.

For the discontinuous case, the methodology outlined in \citep{Verma:2025vkk} is followed: the two phases are separately constructed using the speed-of-sound interpolation method and then stitched together, ensuring thermodynamic consistency via a transition jump (\(\Delta n\)). A check is made to see whether the central density of the $M_{\rm TOV}$ exceeds the PT density. If it does, we identify a discontinuous EoS, as the PT lies within the region probed by the $M_{\rm TOV}$. 

\paragraph{Inference Framework}
The inference framework adopted in this work is based on the methodology developed in \textit{CompactObject}~\citep{Huang_2024rfg}. 
A systematic Bayesian analysis of the posterior distributions for each of the three EoS categories using different combinations of neutron star observational constraints is performed.  
A pair of parameters characterise each segment: the chemical potential $\mu_i$ and the squared speed of sound $c_{s,i}^2$. Consequently, the model contains 10 free parameters. These parameters are inferred within a Bayesian inference framework. This approach differs from earlier studies (e.g., Ref.~\citep{Altiparmak:2022bke, Annala:2019puf}) in which millions of EoS samples were generated via uniform sampling and subsequently filtered to retain only those satisfying all theoretical and observational constraints. As a result, the total number of posterior samples obtained in this analysis is not directly comparable to that obtained using such rejection-sampling methods. Unlike rejection-sampling methods, our approach directly explores the posterior distribution, yielding more efficient, probabilistically rigorous inference of the EoS parameters. For each EoS category, the posterior consists of approximately $5 \times 10^3$ samples. These samples are obtained by drawing approximately $2 \times 10^6$ likelihood evaluations during the inference process using the \texttt{UltraNest} nested sampling algorithm \citep{buchner2021ultranestrobustgeneral}. We incorporate constraints from gravitational-wave observations and X-ray pulse-profile modelling through dedicated likelihood functions. These datasets are treated as statistically independent and are evaluated within a Bayesian framework.

\paragraph{Gravitational-wave likelihood.}
Constraints from the binary neutron star merger GW170817 are implemented using publicly available posterior samples of the component masses and tidal deformabilities~\citep{LIGOScientific:2018hze}. A kernel density estimation (KDE) technique is employed to reconstruct the joint probability distribution
$P(\mathcal{D}_{\mathrm{GW}} \mid m_1, m_2, \Lambda_1, \Lambda_2)$
from the posterior samples.

For a given equation of state parameter set $\theta$, the tidal deformabilities $\Lambda_1$ and $\Lambda_2$ are computed as functions of the component masses. The resulting likelihood is obtained by marginalising over the binary component masses,
\begin{align}
\mathcal{L}^{\mathrm{GW}} = 
\int_{M_l}^{M_u} \mathrm{d}m_1 
\int_{M_l}^{m_1} \mathrm{d}m_2\,
P(m_1 \mid \theta)\,
P(m_2 \mid \theta)\, \nonumber \\
P\!\left(\mathcal{D}_{\mathrm{GW}} \mid 
m_1, m_2, 
\Lambda_1(m_1,\theta), 
\Lambda_2(m_2,\theta)\right),
\label{eq:L_GW_rewrite}
\end{align}
where the prior on the neutron star mass is assumed to be uniform,
\begin{equation}
P(m \mid \theta) =
\begin{cases}
\dfrac{1}{M_u - M_l}, & M_l \le m \le M_u, \\
0, & \text{otherwise},
\end{cases}
\end{equation}
with the lower bound fixed at $M_l = 1\,M_\odot$ and the upper bound set by the maximum mass $M_u = M_{\max}(\theta)$ supported by the equation of state.

\paragraph{NICER likelihood:}
We also incorporate constraints from NICER observations, which provide joint posterior distributions for the neutron star's mass and radius. 
The NICER likelihood is then evaluated by marginalising over the neutron star mass,
\begin{equation}
\mathcal{L}^{\mathrm{NICER}} =
\int_{M_l}^{M_u} \mathrm{d}m\,
P(m \mid \theta)\,
P\!\left(\mathcal{D}_{\mathrm{NICER}} \mid m, R(m,\theta)\right).
\label{eq:L_NICER_rewrite}
\end{equation}

This procedure is applied independently to the NICER measurements of PSR~J0030+0451~\citep{Riley:2019yda}, PSR~J0740+6620~\citep{Riley:2021pdl}, PSR~J0437$-$4715~\citep{Choudhury:2024xbk}, and PSR~J0614$-$3329~\citep{Mauviard:2025dmd}. The resulting likelihoods are combined in different configurations depending on the specific dataset selection adopted for each result set.

\paragraph{CET likelihood.}

For a given EoS parameter set $\theta_{\rm EoS}$, the likelihood contribution from the CET region is modelled as a Gaussian,
\begin{equation}
\mathcal{L}_{\rm CET}(\theta_{\rm EoS})
=
\frac{1}{\sqrt{2\pi}\,\sigma}
\exp\left[
-\frac{1}{2}
\left(
\frac{P(\theta_{\rm EoS}) - P}
{\sigma}
\right)^{2}
\right],
\end{equation}
where $P(\theta_{\rm EoS})$ is the pressure predicted by the EoS, evaluated at the same reference energy densities as the CET data through interpolation of the EoS. Here, $P$ denotes the mean CET pressure value at a given energy density, while $\sigma$ represents the corresponding deviation of the CET data. Both $P$ and $\sigma$ are obtained directly from the tabulated CET limits of \citep{Annala2020}

\section{Results}

The results are presented for three EoS classes, incorporating the observational constraints of PSR J0030+0451, PSR J0740+6620, PSR J0437-4715, and the recent PSR J0614-3329, as well as the GW170817 constraints. To better understand the constraining property of PSR J0614-3329 and GW170817, we have performed inference runs with and without these observations. All the NICER constraints, except those for PSR J0614-3329, are applied first and are named PSR. Further, one applies the GW170817 constraint (PSR+GW) and the PSR J0614-3329 constraint (PSR+J0614) separately. Finally, we present our findings after imposing bounds from both NICER and GW170817 (PSR+J0614+GW). 

\begin{figure*}
    \centering
    \includegraphics[scale = 0.3]{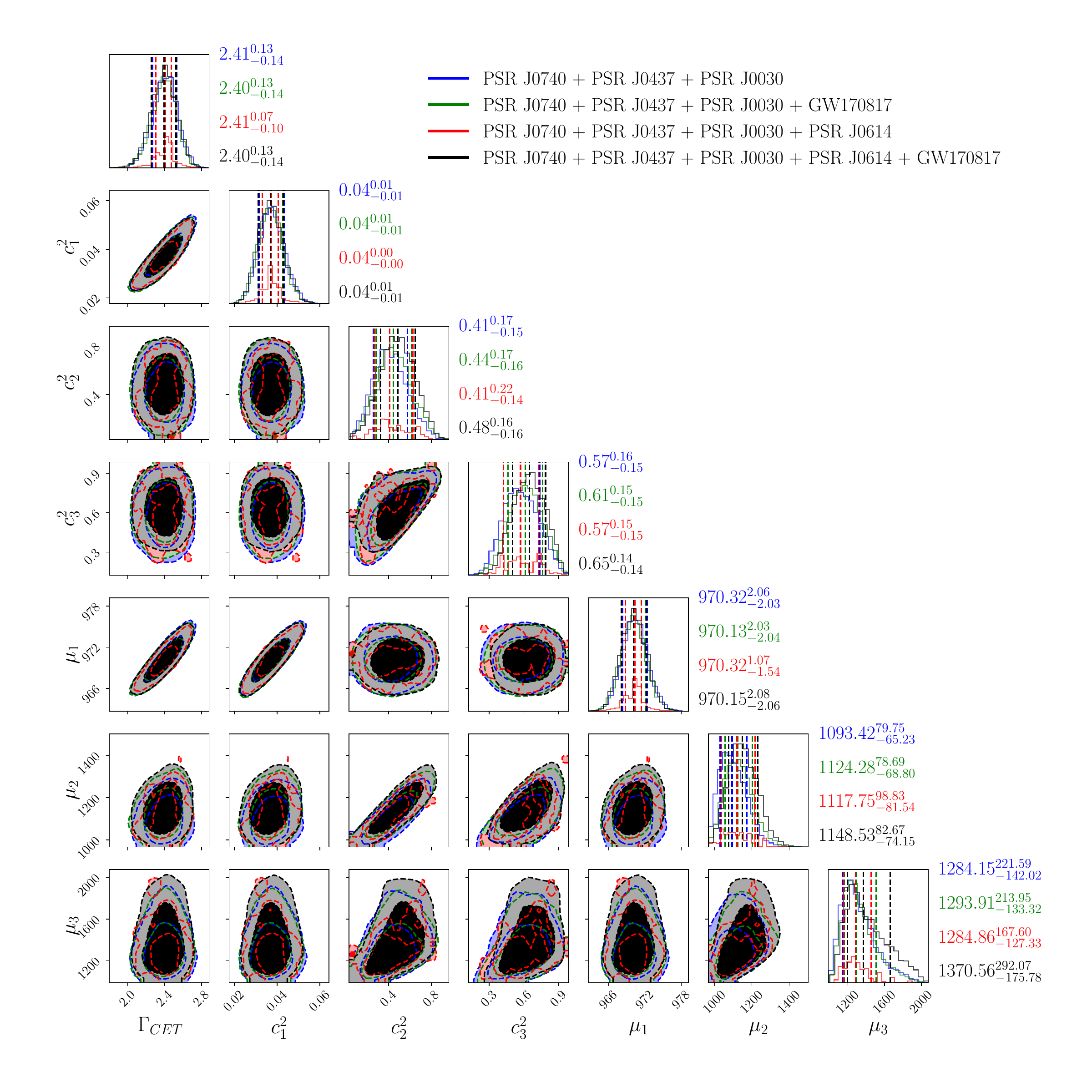}
    \caption{Posterior distribution of EoS construction parameters \{\(c_s^2, \mu\)\} for the monotonic set of EoS is shown in the figure. Different contour marks different sequential observations. The histogram shows the marginalised posterior density of the parameter for the given observation constraint, along with the best-fit values, which are shown in different colours.}
    \label{Fig:mono params}
\end{figure*}

\begin{figure*}
    \centering
    \includegraphics[scale = 0.28]{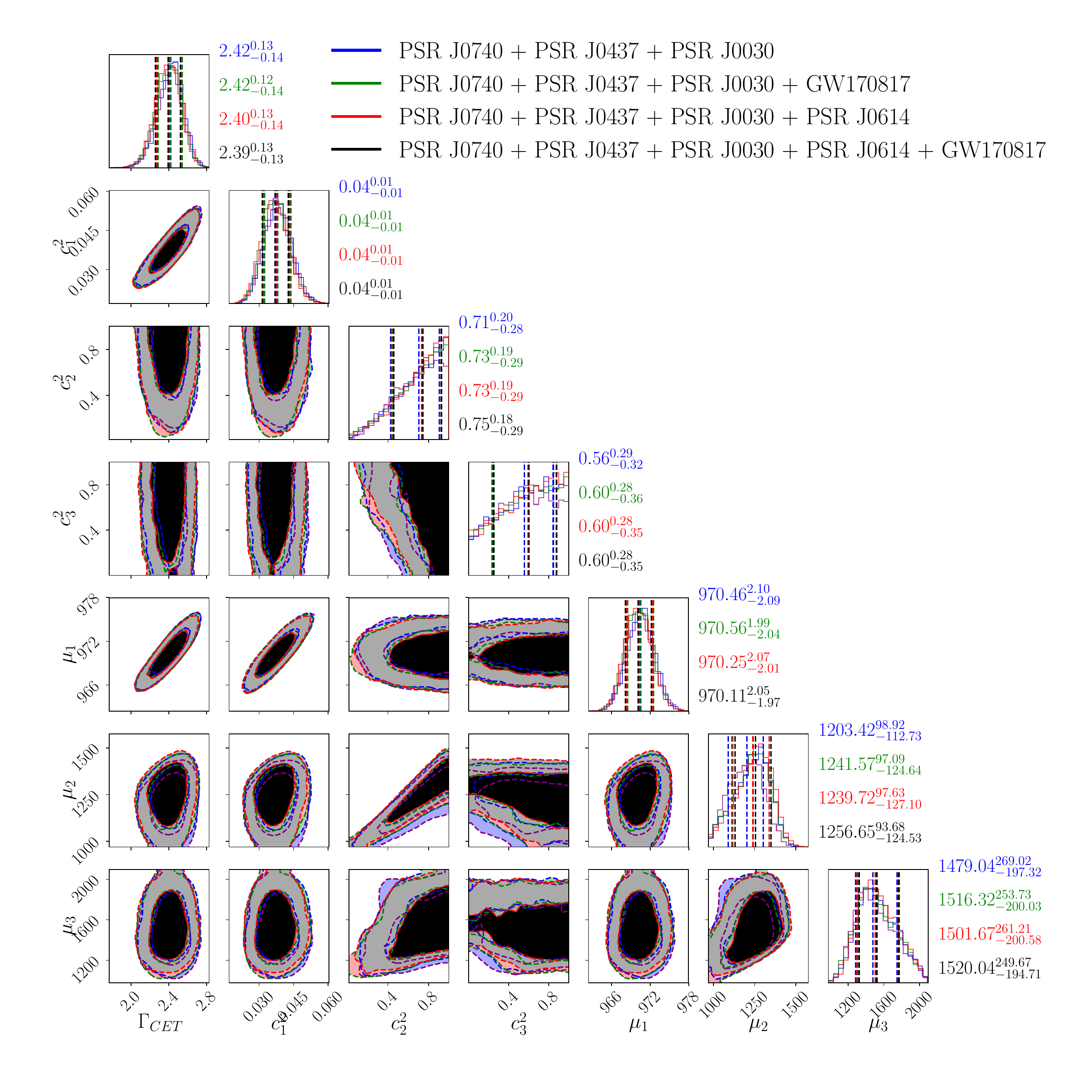}
    \caption{Posterior distribution of EoS construction parameters \{\(c_s^2, \mu\)\} for the non-monotonic set of EoS is plotted in the figure. The nomenclature of the figure remains the same as that explained in Fig. \ref{Fig:mono params}.}
    \label{Fig:Non-mono params}
\end{figure*}

\begin{figure*}
    \centering
    \includegraphics[scale = 0.26]{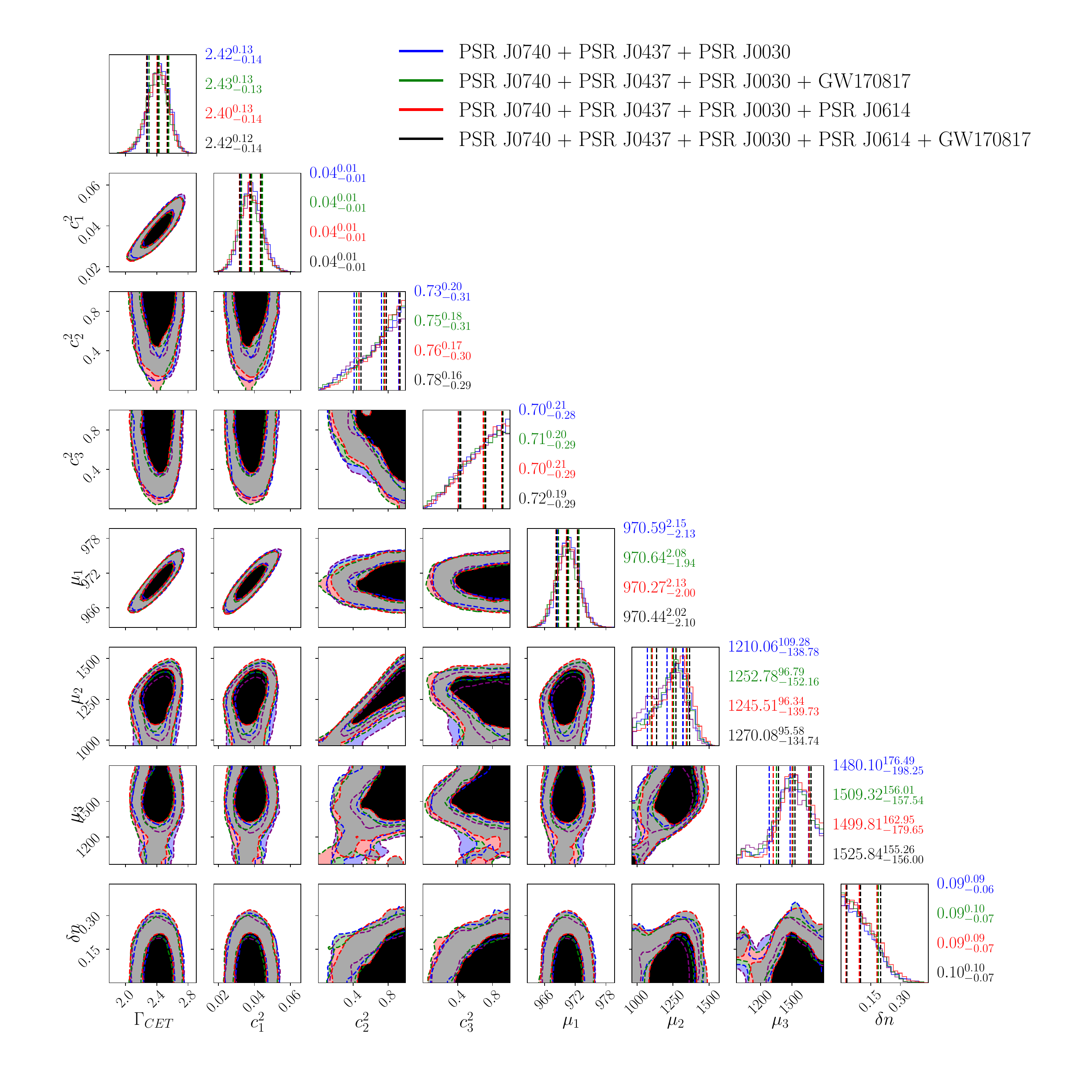}
    \caption{Posterior distribution of EoS construction parameters \{\(c_s^2, \mu\)\} for the discontinuous set of EoS is shown in the figure. The nomenclature remains the same.}
    \label{Fig:discontinuous params}
\end{figure*}

\subsection*{The posterior distributions of the EoS Parameters}

The resulting posterior distributions of the EoS construction parameters $\{c^2_s, \mu\}$ for the three classes of EoS are presented in \ref{Fig:mono params}, Figs.~\ref{Fig:Non-mono params}, and \ref{Fig:discontinuous params} for the monotonic, non-monotonic, and discontinuous sets, respectively. These corner plots display the marginalised one-dimensional and two-dimensional posterior distributions for four combinations of observational constraints: PSR J0740, PSR J0437, PSR J0030 (blue), with GW170817 added (green), PSR J0614-3329 added (red), and all constraints combined (black).

The first parameter, $\Gamma_{CET}$, is the adiabatic index of the polytropic extension that connects the low-density crust EoS to the first segment of the speed of sound parameterisation. This region of the EoS, spanning from $0.5\,n_0$ to $1.1\,n_0$, is constrained by CET calculations \citep{Hebeler_2013}. The remaining parameters consist of five pairs of squared speed of sound values ($c^2_{s,1}$, $c^2_{s,2}$, $c^2_{s,3}$, $c^2_{s,4}$, $c^2_{s,5}$) and their corresponding chemical potentials ($\mu_1$, $\mu_2$, $\mu_3$, $\mu_4$, $\mu_5$), which allows for the piecewise-linear interpolation of $c^2_s(\mu)$ from the CET matching point up to the pQCD regime. It should be noted that Figs.\ref{Fig:mono params}, \ref{Fig:Non-mono params},\ref{Fig:discontinuous params} omits the parameters of $\{c^2_{s,4},c^2_{s,5},\mu_{4},\mu_{5}\}$ as they lie above the chemical potentials realised by our ensemble of stellar models and hence are poorly constrained by the astrophysical observations. For the discontinuous class, an additional parameter $\delta n$ characterises the density jump at the PT. The prior ranges considered for all parameters correspond to uniform priors bounded as:
\begin{itemize}
    \item $\Gamma_{\rm CET} \in \mathcal{U}(1.77, 3.23)$
    \item $c^2_{s,i} \in \mathcal{U}(0, 1)$ for $i = 1, 2, 3,4,5$ (constrained by causality)
    \item $\mu_i \in \mathcal{U}(\mu_{\rm CET}, 2.6\,\text{GeV})$ for $i = 1, 2, 3, 4, 5$
    \item $\delta n \in \mathcal{U}(0, 0.8 fm^{-3})$ (for discontinuous class only)
\end{itemize}

For all three classes, the $\Gamma_{\rm CET}$ also remains well-constrained across all observational combinations because of CET constraints. The first segment of the squared speed of sound, $c^2_{s,1}$, is tightly constrained to approximately $0.04 \pm 0.01$, as this parameter probes densities just above the CET regime. The higher segments $c^2_{s,2}$ and $c^2_{s,3}$ exhibit broader distributions, reflecting the greater uncertainty at higher densities. For the monotonic class (Fig.~\ref{Fig:mono params}), $c^2_{s,2}$ clusters around $0.41$--$0.46$ and $c^2_{s,3}$ around $0.57$--$0.62$, with the central values exhibiting monotonically increasing behavior ($c^2_{s,2} < c^2_{s,3}$). In contrast, for the non-monotonic class (Fig.~\ref{Fig:Non-mono params}), $c^2_{s,2}$ and $c^2_{s,3}$ have median values around $0.71$--$0.75$ and $0.56$--$0.60$, respectively, reflecting the characteristic rise and subsequent fall in the speed of sound. In Fig.~\ref{Fig:discontinuous params}, the density jump parameter $\delta n$ remains relatively small, centred around $0.09$--$0.10$, indicating that the observational constraints prefer modest density discontinuities.

Across all three classes, several common trends emerge from the Bayesian analysis. First, the low-density behaviour characterised by $\Gamma_{\rm CET}$ and $c^2_{s,1}$ is robustly constrained across EoS classes and remains largely independent of the high-density observational constraints. Second, the addition of PSR J0614-3329 and GW170817 systematically shifts the chemical potential parameters toward higher values across all EoS classes. Third, the intermediate and high-density speed of sound parameters ($c^2_{s,2}$, $c^2_{s,3}$) show broader posterior distributions, reflecting the remaining uncertainty in the EoS at densities beyond a few times nuclear saturation density.

\begin{table*}%[h]
    \centering
    \begin{tabular}{l l l l l l}
        \toprule
        Class: Monotonic & ln Z & $R_{1.4}$ (km) & $\epsilon_{TOV}(MeV/fm^3)$ & $\Lambda_{1.4}$ & $\mathcal{C}_{max}$\\ 
        \midrule\midrule
        PSR & -0.6268 & $11.42-13.04$ & $1263 \pm 159$ & $261-717$ & $0.3380$\\
        PSR + J0614 & -1.4038 & $11.24-12.51$ & $1355\pm 140$ & $231-514$ & $0.3337$\\
        PSR + GW & -8.5769 & $11.31-12.56$ & $1353\pm151$ & $232-530$ & $0.3325$\\
        PSR + J0614 + GW & -9.0486 & $11.10-12.31$ & $1423 \pm 139$ & $215-465$ & $0.3323$\\
        \bottomrule
    \end{tabular}
    \caption{Summary of monotonic EoS properties and derived stellar characteristics from the full Bayesian run performed under different observational constraints.}
    \label{Tab:Mono}
\end{table*}

\begin{table*}%[h]
    \centering
    \begin{tabular}{l l l l l l}
        \toprule
        Class: Non-monotonic & ln Z & $R_{1.4}$ (km) & $\epsilon_{TOV}$ $(MeV/fm^3)$ & $\Lambda_{1.4}$ & $\mathcal{C}_{max}$\\ 
        \midrule\midrule
        
        PSR & -0.6849 & 11.36-12.91 & $1195{\pm}172$ & 248-710 & 0.3291\\
        
        PSR + J0614 & -1.3246 & 11.19-12.38 & $1295{\pm}153$ & 225-476 & 0.3328\\
        
        PSR + GW & -8.4584 & 11.23-12.49 & $1284{\pm}151$ & 231-507 & 0.3288\\
        
        PSR + J0614 + GW & -8.8311 & 11.13-12.24 & $1338{\pm}147$ & 215-435 & 0.3270\\
        \bottomrule
    \end{tabular}
    \caption{Summary of non-monotonic EoS properties and derived stellar characteristics from the full Bayesian run performed under different observational constraints.}
    \label{Tab:Non-Mono}
\end{table*}

\begin{table*}%[h]
    \centering
    \begin{tabular}{l l l l l |l l}
        \toprule
        Class: Discontinuous & ln Z & $R_{1.4}$ (km) & $\epsilon_{TOV}$ $(MeV/fm^3)$ & $\Lambda_{1.4}$ & $\mathcal{C}_{max}$\\ 
        \midrule\midrule
        
        PSR & -1.2086 & 11.28-12.89 & $1265\pm186$ & 264-731 & 0.3243\\
        
        PSR + J0614 & -2.0142 & 11.12-12.40 & \(1275\pm179\) & 223-532 & 0.3241\\
        
        PSR + GW & -9.0327 & 11.16-12.86 & \(1267\pm162\) & 251-504 & 0.3254\\
        
        PSR + J0614 + GW & -9.7069 & 11.07-12.48 & $1321\pm182$ & 238-468 & 0.3244\\
        \bottomrule
    \end{tabular}
    \caption{Summary of Discontinuous EoS properties and derived stellar characteristics from the full Bayesian run performed under different observational constraints.}
    \label{Tab:Discont}
\end{table*}

The Bayesian evidence ($\ln Z$) in the Tables is calculated using our nested sampling approach. It is to be noted that incorporating PSR J0614$-$3329---either with the previous NICER observations or with the combined PSR+GW constraints---does not significantly alter the Bayesian evidence ($\ln Z$). According to the Jeffreys scale \citep{Jeffreys1961}, a difference of $|\Delta \ln Z| < 1$ indicates inconclusive evidence, $1 < |\Delta \ln Z| < 2.5$ suggests weak to moderate evidence, $2.5 < |\Delta \ln Z| < 5$ indicates strong evidence, and $|\Delta \ln Z| > 5$ represents decisive evidence for one model over another. A summarised overview of our results is presented in Table \ref{Tab:Mono}, \ref{Tab:Non-Mono}, and \ref{Tab:Discont} for the monotonous, non-monotonous, and discontinuous sets, respectively. 
The changes in $\ln Z$ observed upon adding PSR J0614$-$3329 fall within the inconclusive to weak range, suggesting that the new data do not strongly favour or disfavour any particular EoS class. Since the Bayesian evidence depends on the integral over the prior volume, and the underlying data differ across inference runs, a direct comparison of $\ln Z$ values is not strictly valid. Nonetheless, $\ln Z$ serves as a helpful diagnostic for assessing the compatibility of the observational data with the assumed model.

\subsection*{M-R constraints}\label{3.1}

\begin{figure*}
    \centering
   \includegraphics[width = 0.33\linewidth]{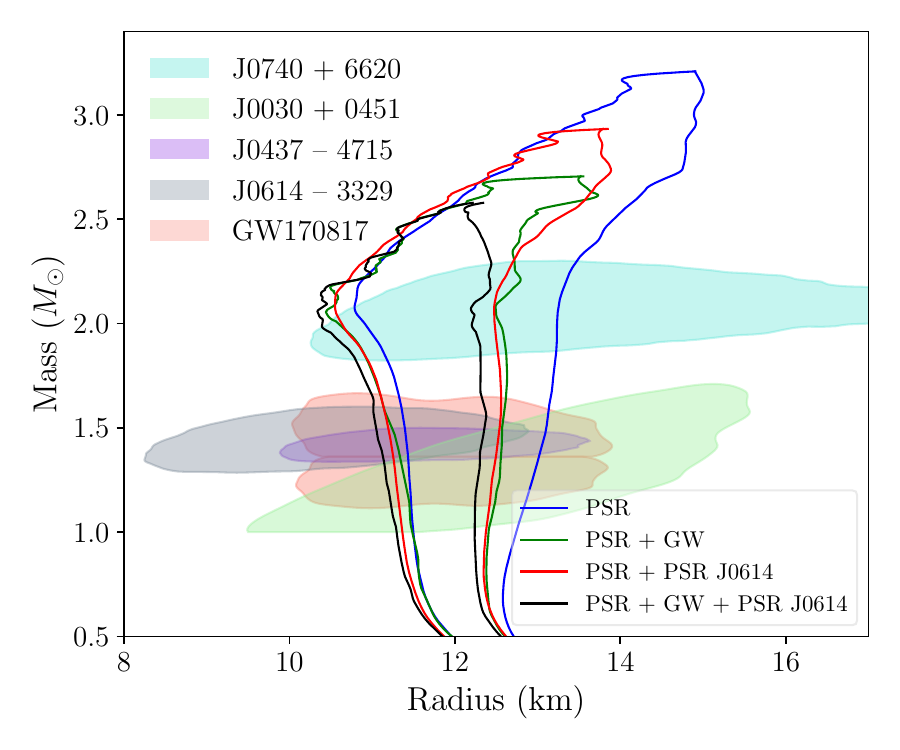}
    \includegraphics[width = 0.33\linewidth]{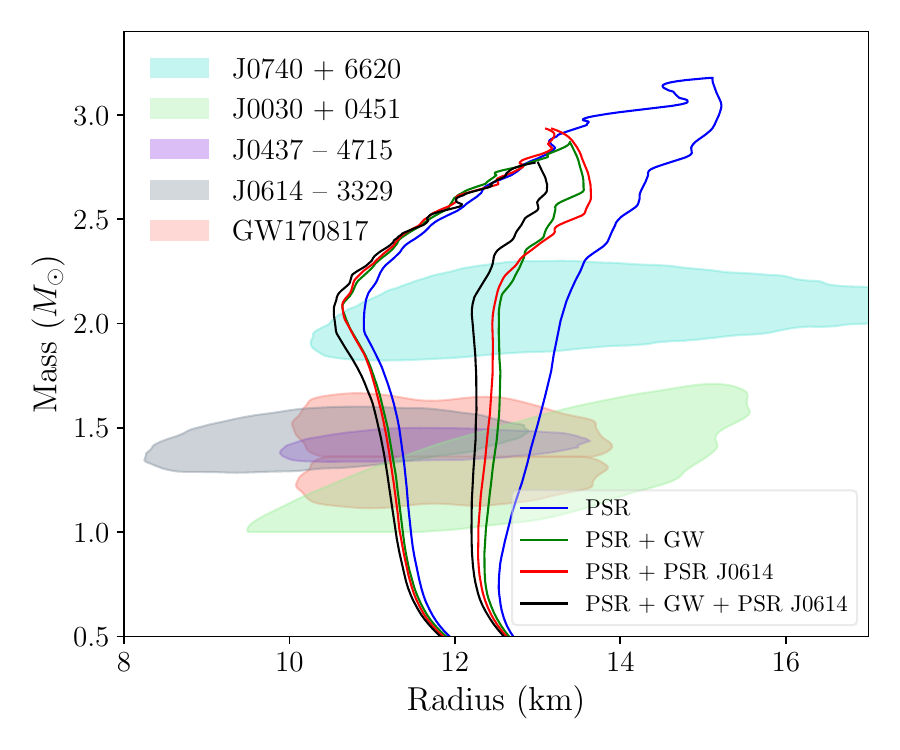}
    \includegraphics[width = 0.33\linewidth]{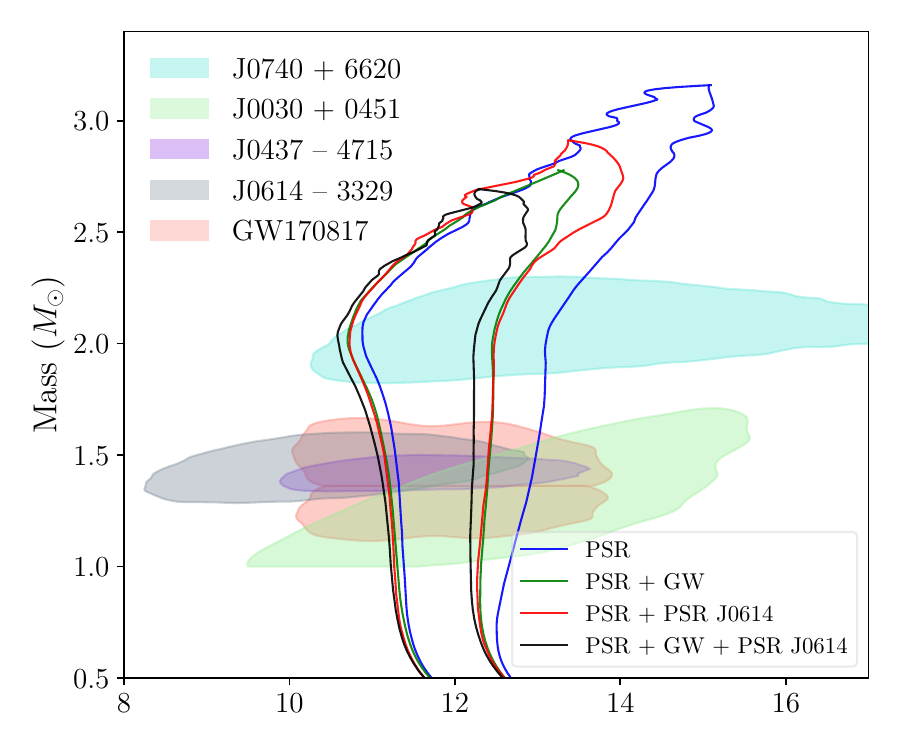}
    \caption{The 90\% CI of M-R posterior distribution for different combinations of observational constraints for the non-monotonic set (left), monotonic set (middle), and discontinuous (right) is shown. Different colour contours signify different observational constraints.}
    \label{fig:MR}
\end{figure*}

In Fig. \ref{fig:MR}, we present the 90\% credible interval (CI) of the M-R posterior distributions for all the sets: monotonic, non-monotonic and discontinuous EoS sets. 
%A consistent trend is observed in the M--R posteriors in all cases. 
Additional bounds from GW170817 and PSR J0614-3329 shift the M--R posterior toward smaller radii, as expected. However, the shift toward smaller radii is slightly more substantial for GW170817. The results are also very evident from the numbers depicted in the Tables. 
Nevertheless, the overall difference between the two observations remains small, primarily because PSR J0740+6620 favours larger-radius stars. This interplay between the preference for larger radii from PSR J0740+6620 and smaller radii from PSR~J0614-3329 and GW170817 leads to a compensation, resulting in the posterior distribution occupying the broadly very similar region of the M--R space as shown in Fig. \ref{fig:MR}. This compensation for the contrasting observations also reduces the back-bending property of the M-R sequences. The back-bending property determines whether an M-R sequence bends backwards at higher masses. Usually, backwards bending of the M-R sequence corresponds to massive stars (the $M_{TOV}$ increases). Another consistent trend observed across all sets is the reduction in maximum mass. This decrease is more pronounced when the GW170817 constraint is applied than when PSR J0614-3329 is included.

\subsection*{Sequences with maximum mass and maximum compactness}
\label{result:compactness}
The NICER and GW constraints systematically favour softer equations of state at intermediate densities, which have profound implications for neutron star structure. To understand these implications, we examine two physically distinct stellar configurations: the maximum-mass and maximum-compactness sequences.

Compactness, defined as $C = GM/Rc^2$, quantifies how relativistic a stellar object is. For a \textit{single} equation of state, the maximum-mass and maximum-compactness configurations coincide---they represent the same star at the TOV limit~\cite{Rezzolla:2025pft}. However, when considering an \textit{ensemble} of equations of state, the EoS producing the largest maximum mass differs fundamentally from the EoS producing the largest maximum compactness. This distinction reveals a remarkable pattern when combined with observational constraints.

\begin{figure*}
    \centering
    \includegraphics[width=0.33\linewidth]{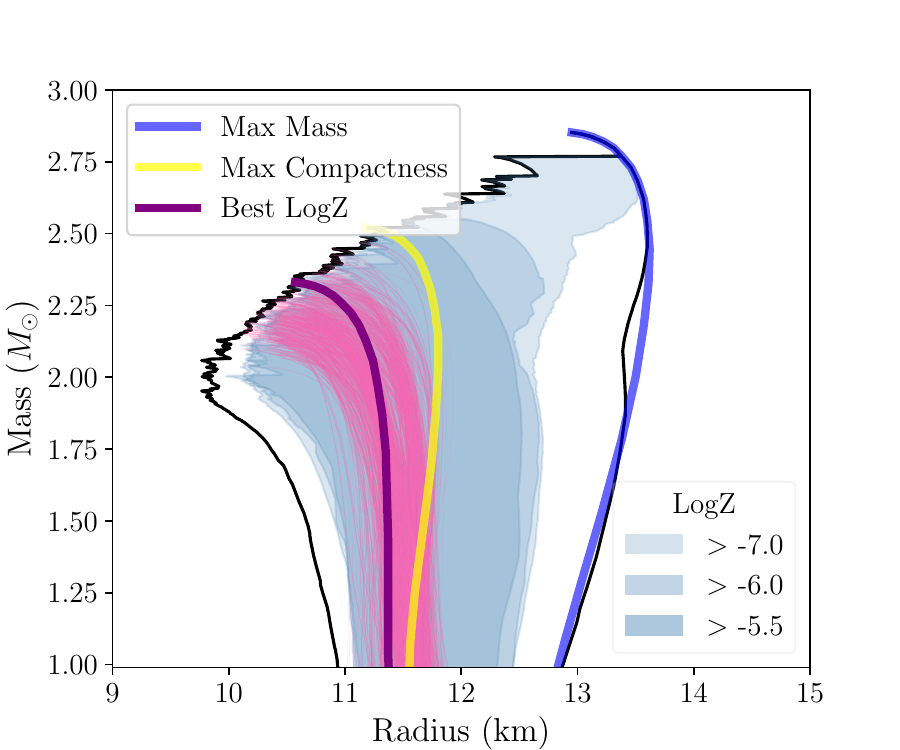}
    \includegraphics[width=0.33\linewidth]{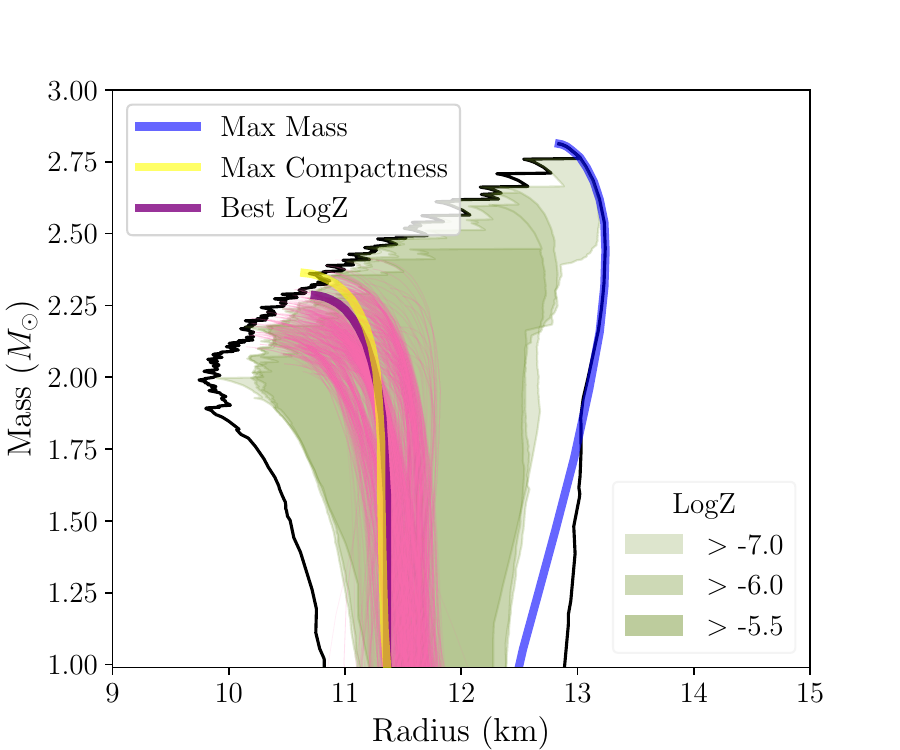}
    \includegraphics[width=0.33\linewidth]{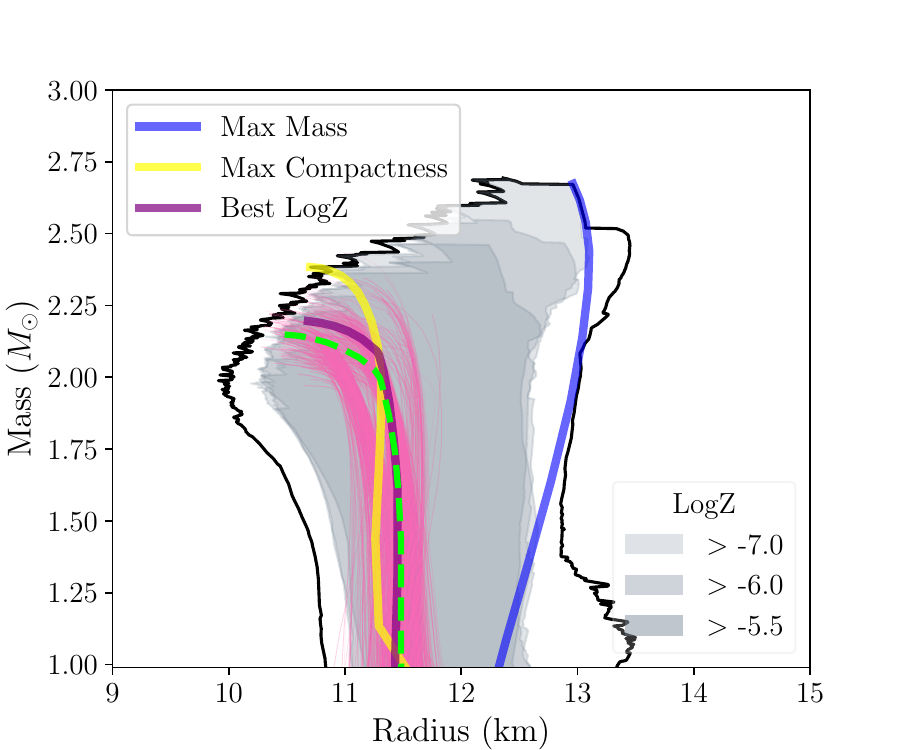}
\caption{Along with the maximum mass and maximum compact M-R sequence, we also plot the most probable (having the highest Ln Z value) EoS for each set. Also plotted the top 500 most probable M-R curves with thin red lines.}
\label{fig:probable}
\end{figure*}

\begin{figure*}
    \centering
    \includegraphics[width=0.33\linewidth]{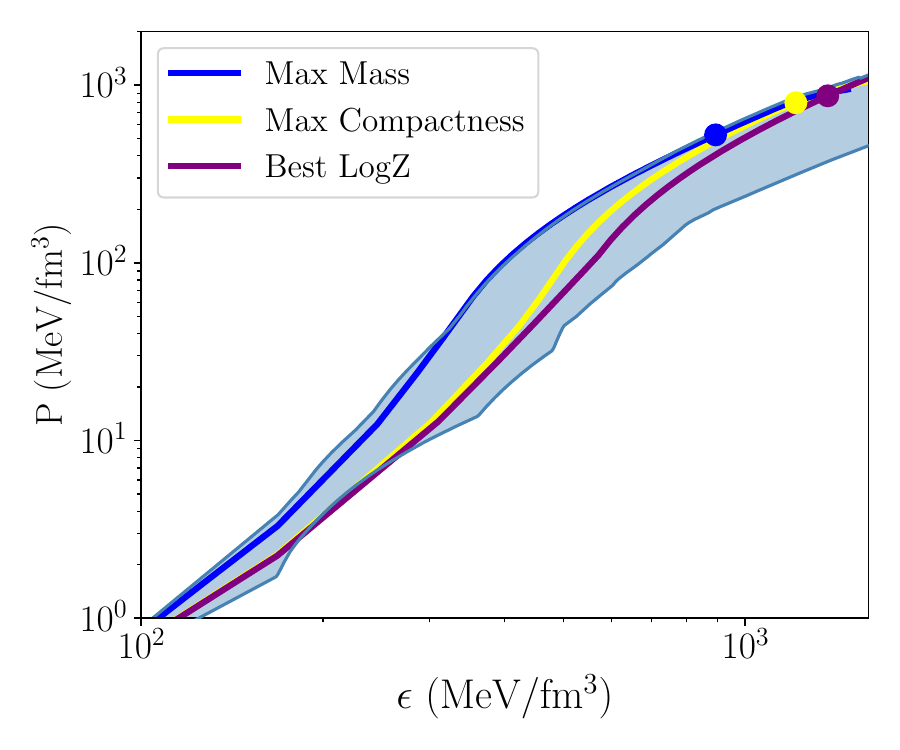}
    \includegraphics[width=0.33\linewidth]{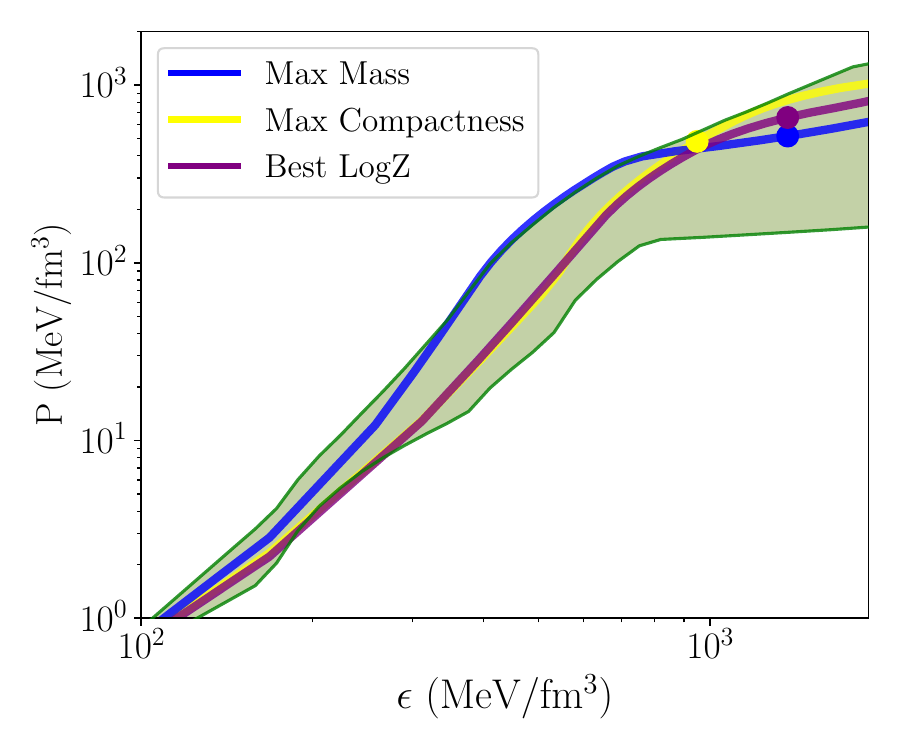}
    \includegraphics[width=0.33\linewidth]{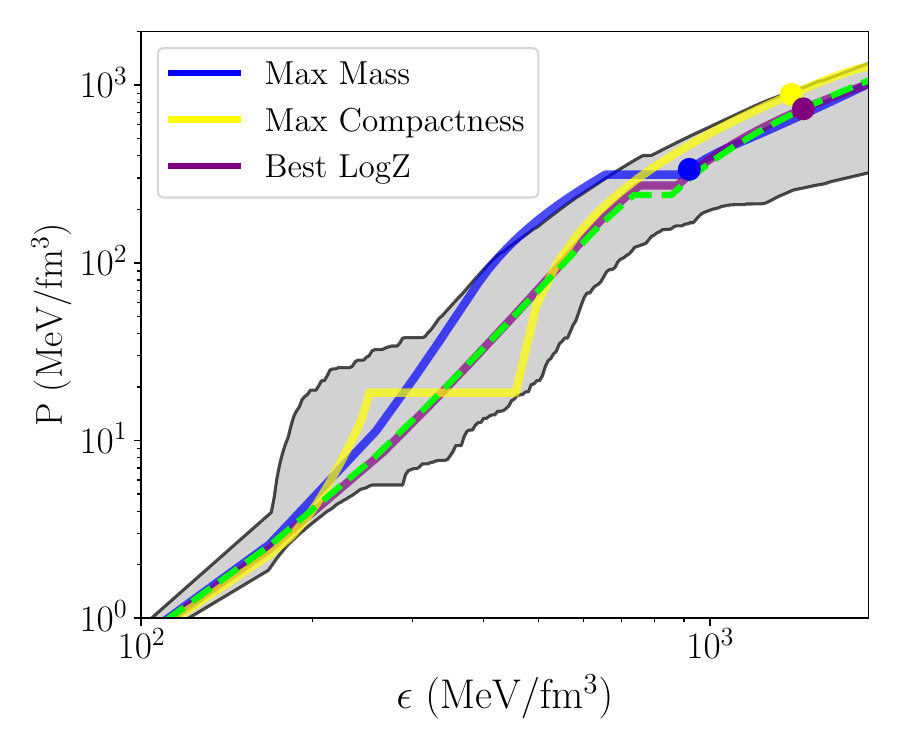}
\caption{The entire span of EoS for the three classes is shown in different panels of the plot. The EoS corresponding to the maximum compactness and maximum mass is shown explicitly with red and blue curves, respectively. The scatter points correspond to the central energy density/pressure of the stellar model for maximum compactness (red) and maximum mass (blue).}
\label{fig:eos-compact}
\end{figure*}

The maximum-mass EoS (blue curves in Figs.~\ref{fig:probable}--\ref{fig:eos-compact}) exhibits pronounced stiffening from low densities onward. This behaviour is intuitive: uniform stiffness provides pressure support against gravitational collapse across all density scales, enabling the star to sustain the largest possible mass. These configurations occupy the extreme right edge of the M--R posterior, at large radii.

The maximum-compactness EoS (red curves) displays qualitatively different behaviour: softness at low-to-intermediate densities followed by pronounced stiffening at high densities. This structure optimises compactness by simultaneously minimising the radius (through weak envelope pressure support) while maintaining high mass (through strong core pressure support). The maximum-compactness EoS actually exceeds the maximum-mass EoS in pressure at high densities, despite lying below it at intermediate densities---a crossover that reflects this delicate balance.

The central result of this analysis emerges from comparing these sequences with the \textit{most probable} EoS dictated by observations -- that is, the sequence corresponding to the highest Bayesian evidence ($\ln \mathcal{Z}$) (magenta curves in Figs.~\ref{fig:probable}--\ref{fig:eos-compact}). The observationally favoured EoS shares the same qualitative structure as the maximum-compactness EoS: relative softness at intermediate densities transitioning to stiffness at high densities. This convergence is not coincidental: Together with the NICER and GW170817 observations, the CET constraints independently and consistently push the allowed EoS parameter space toward configurations that gravity naturally favours —compactification — thereby approaching higher compactness.

This finding suggests that nature may be selecting for neutron stars near their compactness limits. The observational preference for softer intermediate-density EoS, combined with the requirement to support $\sim 2\,M_\odot$ stars, naturally produces stellar configurations that are more compact than what uniformly stiff EoS would predict. Rather than being an artefact of limited data, this convergence between observational constraints and maximum-compactness configurations points toward a fundamental characteristic of dense matter: the equation of state likely undergoes significant softening at intermediate densities---potentially signalling the onset of new degrees of freedom---before stiffening again to support the most massive observed neutron stars.

\textbf{Phase transition effects:} For the discontinuous EoS class, the distinction between maximum-mass and maximum-compactness sequences has a clear interpretation in terms of the phase transition density:
\begin{itemize}
    \item The \textbf{maximum-mass configuration} corresponds to an EoS with a phase transition occurring at high density. The resulting star has only a small quark core (or no quark core at all), with most of the stellar interior composed of hadronic matter.
    \item The \textbf{maximum-compactness configuration} corresponds to an EoS with a phase transition occurring at a lower density. This produces a star with a substantial quark core. Since quark matter is typically softer than hadronic matter, it is more easily compressed, leading to a smaller radius and hence greater compactness.
\end{itemize}

The discontinuous class of EoS have more freedom, and therefore, the maximum mass and maximum compact sequence are governed by the transition pressure (or density) along with the width of the discontinuity. Higher compactification is achieved by a softer EoS at low density, suggesting that PT occurs at low density (occurring at $1.5$ times the saturation density). Thereafter, the EoS stiffens significantly at high density and achieves a maximum compactness. However, this is a marginal case, not completely ruled out but very unlikely, as the Bayesian predictions favour late PT and a small density discontinuity. Therefore, for the discontinuous EoS, the maximum compact EoS and most probable EoS differ considerably. However, if one rules out very early PT ($< 3$ times the saturation density), the maximum compact EoS is again very similar to that of the most probable EoS (as shown in Fig \ref{fig:probable} -- \ref{fig:eos-compact}).

This convergence is observed across all three EoS classes, although it is most pronounced for the non-monotonic and discontinuous cases. For EoS classes featuring phase transitions, this result has an intuitive explanation: The observational constraints prefer softer EoS, and the softening associated with the phase transition facilitates gravitational compression, naturally driving the stellar structure toward higher compactness. If the assertion is correct, one expects the maximum mass of NSs to decrease significantly, and depending on the EoS class, it can range from $ 2.1$ to $2.4$ solar masses.

\section{Summary and Conclusion}

NS physics depends heavily on the properties of matter at high density, which is found in the cores of NSs. The astrophysical observations from NICER and GW have constrained the matter properties in NS cores to a large extent. At high density, the expected degrees of freedom are quarks rather than hadrons; however, the density at which PT occurs is still debated. Given the uncertainties, one can model matter at high densities into three distinct classes: the monotonous (hadronic) EoS, the non-monotonous (smooth PT) EoS, and the discontinuous EoS with a density discontinuity. 

The Bayesian analysis of the parameters (for the three classes of EoS) shows that the low-density EoS remains well-constrained till $2$ times the nuclear saturation density, independent of its high-density behaviour. In contrast, the high-density EoS shows a broader distribution, reflecting the uncertainty of the EoS. The non-monotonous EoS shows a peak in the intermediate density range, and the observational constraints favour a small density discontinuity for the discontinuous class.
The imposition of observational constraints shrinks the M-R contour mostly from the right, rejecting very massive NSs. The astrophysical constraints seem to favour late PT for both non-monotonic and discontinuous EoS, confining quark matter to the cores of sufficiently massive stars. 

Given an EoS, the maximum mass and maximum compact configuration coincide as one moves along the M-R sequence just prior to the gravitational instability. However, for an ensemble of EoS, the EoS which yields the maximum mass configuration is not the same EoS that generates the maximum compact configuration. The maximum mass EoS shows considerable stiffening from low density to sufficiently high density, providing high-pressure support against gravitational collapse and yielding the maximum possible mass for the NS. In contrast, the maximum compact EoS have a softer EoS at low density to minimise the radius, followed by pronounced stiffening at high density to support high mass, thereby producing maximum compact NS. For the discontinuous class, the EoS that generates the maximum mass has PT at high density, resulting in a massive NS with a small quark core. In contrast, the EoS yielding the maximum compact star has PT at a lower density, resulting in a very compact NS with a substantially large quark core. The most striking feature of our analysis is the closeness of the most probable M-R sequences to the maximum compact sequence. The convergence is striking and points towards a fundamental aspect of dense matter, which softens at intermediate densities---signalling the onset of new degrees of freedom---before stiffening again to support the most massive neutron stars possible. The outcome is also what gravity naturally favours — compactification.

The discontinuous EoS with an unlikely low-density PT is expected to unravel in the near future, as the $\chi$CET band is expected to extend further from the nuclear saturation density, or through a NICER measurement of a sufficiently low-mass pulsar. Better constraints on higher-mass pulsars would also provide tighter constraints on the high-density EoS in their cores.
Future GW and/or NICER observations of pulsar mass and radius are expected to provide more stringent constraints and are likely to provide us with a clearer picture of the stellar configuration and the matter properties associated with it. 

\section*{Acknowledgements}

The authors AKS and RM thank the Indian Institute of Science Education and Research Bhopal for providing all the research and
infrastructure facilities. RM acknowledges the Science and Engineering Research Board (SERB), Govt. of India, for financial support in the form of a Core Research Grant (CRG/2022/000663). The simulations were conducted at the Bhaskara HPC in IISER Bhopal. T.M. and R.M. express gratitude for the Deucalion HPC platform in
Portugal, appreciating its resources and technical support,
facilitated by the FCT project 2025.00067.CPCA.A3. T.M.
received support from Fundação para a Ciência e a Tecnologia
(FCT), I.P., Portugal, under the projects UIDB/04564/2020
(doi:10.54499/UIDB/04564/2020), UIDP/04564/2020 (doi:10.
54499/UIDP/04564/2020).

\section*{Data availability statement}

This is a theoretical work and does not include any additional data.

\bibliographystyle{unsrtnat}
\bibliography{References}

\appendix

\section{The likelihood of different M-R sequences for 3 classes of EoS}\label{a-3}

\begin{table}[h!]
\centering
\caption{Likelihood values of different sequences for 3 classes of EoS}
\begin{tabular}{lccc}
\hline
Sequences & Monotonic & Non-monotonic & Discontinuous \\
\hline
Best LogZ & -5.168 & -5.117 & -5.330 \\
Max Compactness  & -6.062 & -6.019 & -5.440 \\
Max Mass  & -12.639 & -11.207 & -13.892 \\
\hline
\end{tabular}
\end{table}

\label{lastpage}
\end{document}